\documentstyle[amssymb,aps,epsfig,twocolumn]{revtex}
\begin{document}
% !!!!!!!!!!!!!!!!!!( PAPER CODE: R4A1550 )!!!!!!!!!!!!!!!!!
\title{{\bf Bose-Einstein condensation of nonzero-center-of-mass-momentum
Cooper
pairs}}
\author{J. Batle, M. Casas}
\address{Departament de F\'{\i }sica, Universitat de les Illes Balears\\
07071 Palma de Mallorca, Spain}
\author{M. Fortes, M.A. Sol\'{\i }s}
\address{Instituto de F\'{\i }sica, Universidad Nacional Aut\'{o}noma de
M\'{e}xico \\
Apdo. Postal 20-364, 01000 M\'{e}xico, DF, Mexico}
\author{M. de Llano, A.A. Valladares}
\address{Instituto de Investigaciones en Materiales, Universidad Nacional
Aut\'{o}noma de M\'{e}xico \\
Apdo. Postal 70-360, 04510 M\'{e}xico, DF, Mexico}
\author{O. Rojo}
\address{PESTIC, Secretar\'{\i }a Acad\'{e}mica \& CINVESTAV - IPN, 04430
M\'{e}xico, DF, Mexico \\
{\small Received (January 15, 2001)}}
%\vspace{-0.5cm}
\maketitle
\noindent{{\bf Abstract}}
\begin{abstract}
Cooper pair (CP) binding with both zero and nonzero center-of-mass momenta
(CMM) is studied with a set of renormalized equations
assuming a short-ranged (attractive) pairwise interfermion interaction.
Expanding the associated dispersion relation in 2D in powers
of the CMM, in weak-to-moderate coupling a term {\it linear} in the CMM
dominates the pair excitation energy, while the quadratic behavior usually assumed in
Bose-Einstein (BE)-condensation studies prevails for any coupling {\it only} in the limit of
zero Fermi velocity when the Fermi sea disappears, i.e., in vacuum. In 3D this same behavior is observed numerically. The linear term, moreover, 
exhibits CP breakup beyond a threshold CMM value which vanishes with coupling.  This makes all the excited (nonzero-CMM) BE levels with preformed CPs collapse into a single ground level so that a BCS condensate (where only zero CMM CPs are usually allowed) appears in zero coupling to be a special case in either 2D or 3D of the BE condensate of
linear-dispersion-relation CPs.

\smallskip



\end{abstract}

%\smallskip
\noindent {\bf 1. Introduction}
\smallskip
\smallskip

We consider an s-wave short-range, attractive (rank-one) separable 
interfermionic potential \cite {SR&N} in $d$-dimensional momentum space  
$V_{pq}=-(v_{0}/L^d)g_{p}g_{q}$ 
where $v_{0}\geqslant 0$ is the interaction strength, $L$ the size of the 
system, and the
$g_{p}$'s are dimensionless form factors of the type
$g_{_{p}}=(1+p^{2}/p_{0}^{2})^{-1/2}$
in which $p_{0}$ is the inverse range of the potential. Thus, e.g.,  $%
p_{0}\rightarrow \infty $ implies $g_{_{p}}=1$ which corresponds to a
contact or delta potential $-v_{0}\delta (\bf r)$ in configuration space. In 
either 2D or 3D such a potential well has an infinite number of bound states. As a result a many-fermion system with this interfermion interaction will collapse in the thermodynamic limit to infinite binding per particle and infinite density.
However, the potential can be \lq \lq regularized," i.e., constructed \cite {GT} 
with $v_0$ infinitesimally small so that it supports a {\it single} bound state.

The CP equation \cite {coop} for two interacting electrons of mass $m$ above 
the Fermi surface, with momenta
wavevectors ${\bf k}_{1}$ and ${\bf k}_{2}$ and finite, nonzero
CMM wavevector ${\bf K}\equiv {\bf k}_{1}+{\bf k}%
_{2}$, and relative
momentum wavevector ${\bf k}\equiv \frac{1}{2}({\bf k}_{1}-{\bf k}_{2})$, gives
the total pair energy $E_{K}\equiv 2E_{F}-\Delta _{K}$ in terms of $v_{0}$,
with
$E_{F}\equiv \hbar^{2}k_{F}^2/2m$ the Fermi energy.  Here
$\Delta _{K}\geqslant 0$ is the CP binding energy; it should not be
confused with
the BCS energy gap $\Delta$. One can eliminate the variable $v_{0}$ in favor, 
in 2D, of the
vacuum bound-state energy $B_{2}\geqslant 0$ of the potential by combining
\cite{PRB} the CP equation with the respective Lippmann-Schwinger one for the 
same interfermion interaction acting not in the Fermi sea but in vacuum. Then
$\Delta _{K}$ can be extracted as a function of $B_{2}$ from the resulting 
{\it renormalized CP equation}
\begin{eqnarray}
&&\sum_{k}\frac{g_{k}^{2}}{B_{2}+\hbar ^{2}k^{2}/m}-
\sum_{{k},(|{\bf K}/2%
\pm {\bf k}|>k_{F})} \nonumber \\
&&\frac{g_{k}^{2}}{\hbar ^{2}k^{2}/m+\Delta
_{K}-2E_{F}+\hbar ^{2}K^{2}/4m}=0.  \label{8} 
\end{eqnarray}

%\vspace{0.2cm}
\smallskip
\noindent {\bf 2. Cooper pair dispersion relation}
\smallskip
\smallskip

After some algebra one finds the remarkable identity, but only in 2D, that
$\Delta_{0}=B_{2}$, i.e., for an attractive delta interaction (regularized or
not) the vacuum and zero-CMM CP binding energies coincide for {\it all}
coupling.
Using $E_{F}/k_{F}\equiv \hbar
v_{F}/2$ one can expand $\Delta_K$ in powers of $K$ for any coupling 
$B_{2}$ and get
\begin{eqnarray}
&&\varepsilon _{K}\equiv ({\Delta _{0}-\Delta _{K})}=\frac{2}{\pi }\hbar
v_{F}K + \nonumber \\
&&\left[ 1-\left\{ 2-\left( \frac{4}{\pi }\right) ^{2}\right\} \frac{%
E_{F}}{B_{2}}\right] \frac{\hbar ^{2}K^{2}}{2(2m)} + O(K^{3}), \quad (2\mbox{D})
\label{dk2}
\end{eqnarray}
%FIGURE 1
\begin{figure}[htbp]
%\centereps{8cm}{5.25cm}{C:/articulo/honolulu/fig1h1.eps}
%\begin{quotation}
\hspace{-0.5cm}\epsfig{figure=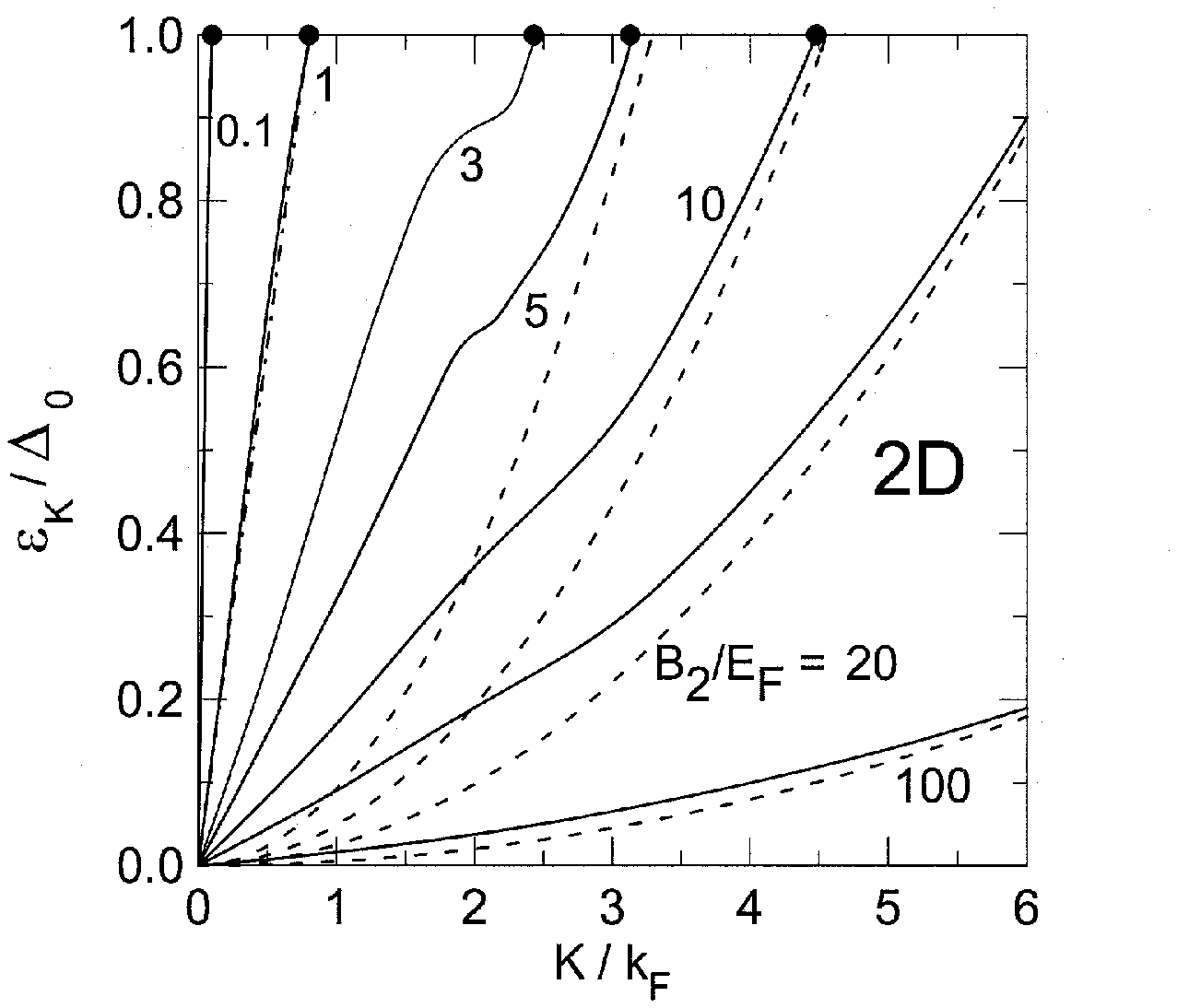,width=8.0cm}
%\seteps{0.0cm}{2.9in}{2.9in}{c:/articulo/honolulu/fig1h1.eps}
\caption[]{
Exact numerical results (full curves) for CP dispersion relation
$\varepsilon _{K} \equiv \Delta _{0}-\Delta _{K}$ (in units of $K=0$ CP binding
energy $\Delta _{0}$) obtained from (\ref {8}) when $g_{k}=1$ for different 
coupling values $B_{2}/E_{F}$. \ CPs break up when $\Delta_{K}$ turns
negative, as indicated by large dots.  The dot-dashed line is the linear
%$2\hbar v_{F}K/{\pi}$ 
approximation (virtually coincident with the exact curve
for all $B_{2} \leq 0.1E_F$)
while the quadratic
%$\hbar^{2}K^{2}/2(2m)$ 
approximation is shown dashed (see text for details).}
\label{Fig 1}
\end{figure}

\noindent where a nonnegative {\it CP excitation energy }$\varepsilon _{K}$ 
has been defined. It is this excitation energy that enters in the BE 
distribution function in determining the critical temperature in a picture of
superconductivity as a BE condensation (BEC) of CPs. The leading term in (\ref{dk2}) is linear in 
CMM, followed by a quadratic term. The latter is precisely the kinetic energy
of what was originally the ordinary CP (and now is what is
sometimes called a ``local pair'')---namely the familiar nonrelativistic
energy of the composite pair of mass $2m$ in
vacuum. This dispersion relation has been
the starting point for virtually all BEC studies of superconductivity (see,
e.g., \cite{SR&N,Blatt,Hauss,Lee,Tolma}, among
others). However, it is clear from (\ref{dk2}) that the quadratic term $\hbar^2K^2/2(2m)$ will 
prevail for any nonzero coupling
{\it only} when $E_{F}/k_{F} \equiv \hbar v_{F}/2 \rightarrow 0$, i.e., in the vacuum limit when there is no Fermi sea.

Figure 1 shows exact numerical results (full curves) of (1) in 2D for different 
$B_{2}/E_{F}$ of the CP excitation energy
$\varepsilon{_{K}/\Delta }_{0}$ as function of CMM $K/k_{F}$.
Note that the CPs {\it break up} at $\varepsilon _{K}/\Delta_{0}=1$ where
$\Delta_{K}=0$, this being marked by large dots in the figure. In addition to the
exact results we also exhibit the linear approximation $2\hbar v_{F}K/{\pi}$
(dot-dashed lines) for small $B_2/E_F$, as well as the quadratic approximation
$\hbar^{2}K^{2}/2(2m)$ (dashed parabolas) for large $B_{2}/E_{F}$.

In 3D one obtains \cite{PC} similar results except that the dimensionless
s-wave
scattering length $k_{F}a$ in vacuum plays the role of a coupling parameter 
instead of the dimensionless binding energy $B_{2}/E_F$ in the 2D case.  Here, the limit
$\Delta_{0}\rightarrow 0$ implies ${a}\rightarrow 0^{-}$ or 
$1/k_F a \rightarrow -\infty $ and corresponds
to weak coupling, while the limit $\Delta_{0}\rightarrow \infty $ implies
${a}\rightarrow 0^{+}$ or $1/k_F a \rightarrow +\infty $ and is strong coupling. 
In fact, for ${a}=-|{a}|\rightarrow 0^{-}$ one finds 
$\Delta_{0}\rightarrow(8E_{F}/e^{2})\exp(-\pi /k_{F}|{a}|)$, a result first 
obtained by Van Hove\cite{vanhove}. On the
other hand, ${a}\rightarrow 0^{+}$ yields $\Delta_{0}
\rightarrow \hbar^2/m {a}^{2}$.  Repeating the expansion carried out in 2D but
without explicitly determining the coefficient of the quadratic term gives
\begin{equation}
\varepsilon {_{K}}\equiv (\Delta _{0}-\Delta _{K})\rightarrow \frac{1}{2}%
\hbar v_{F}K+O(K^{2}), \quad \quad (3 \mbox{D})   \label{dk3}
\end{equation}
i.e., the same result cited in 1964 in Ref. \cite{sch64} for the BCS model
interaction. The linear terms in both Eqs. (\ref{dk2})
and (\ref{dk3}) are identical \cite{physc} for the BCS model interaction in
weak coupling. In this case $g_k = \theta(\hbar^2 k^2/2m- \mbox{max}[0,(E_F-\hbar \omega_D)]) \theta(E_F + \hbar \omega_D - \hbar^2 k^2/2m)$ where $\theta (x)$ is the Heaviside step function and $\omega_D$ the Debye frequency. It becomes $g_k =1$ as $\hbar \omega_D \to \infty $\\

\smallskip
\noindent {\bf 3. Boson number}
\smallskip
\smallskip

Using a statistical model \cite {PA} guaranteeing both thermal and chemical 
equilibrium in an ideal boson-fermion mixture, the number of
bosons $N_{B}(T)$ formed within the $N$-fermion system, valid at and below the
BEC transition temperature $T_c$, is

\begin{eqnarray}
N_{B}(T)&\equiv& {\frac{1}{2}}\left[N-N_0(T) \right] \nonumber \\
&=&{\frac{N}{2}} [ 1-(T/T_{F})%
\ln (1+e^{-\beta \{\Delta_{0}(T)/2-\mu (T)\} } ) ] 
\label{n20t26}
\end{eqnarray}
where $N_0(T)$ is the number of unpaired fermions,  $\Delta_{0}(T)$ the 
appropriate finite-$T$ generalization \cite {PA} of the CP $K=0$ binding energy, $\beta \equiv 1/k_B T$, 
and the ideal Fermi gas chemical potential $\mu (T)$ in 2D is given exactly by 
\begin{equation}
\mu (T)=\beta ^{-1}\ln (e^{\beta E_{F}}-1)\mathrel{\mathop {\longrightarrow
}\limits_{T\rightarrow 0}}E_{F}.  \label{muT}
\end{equation}
\noindent Figure 2 illustrates the zero CMM CP binding energy $\Delta_{0}(T)$ for three values of $B_2 / \mu(T)$.

% FIGURE 2.
\begin{figure}[htbp]
%\hspace{-0.5cm}\epsfig{figure=delta_T.ps,width=8.cm,angle=270}
\hspace{-0.5cm}\epsfig{figure=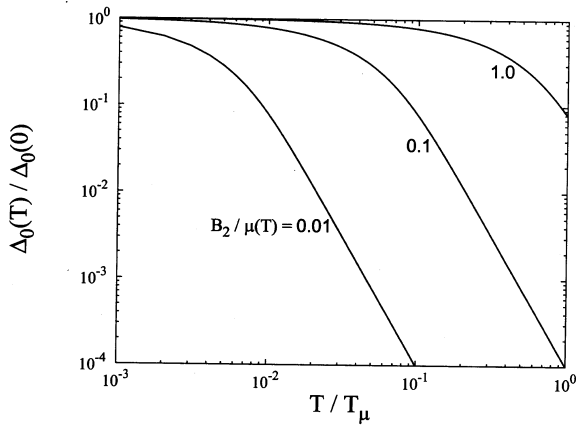,width=8.cm}
\vspace{.5cm}
\caption[]{ Temperature dependence of 2D zero-CMM CP binding energy
$\Delta_{0}(T)$ {\it vs.} $T/T_{\mu}$ for several couplings $B_2/\mu(T)$, where 
$k_B T_{\mu} \equiv \mu$.} 
\label{Fig 2}
\end{figure}

At $T=0$ (\ref {n20t26}) becomes
\begin{eqnarray}
N_B(0)=N\Delta_0(0)/4E_F &\equiv& NB_2/4E_F \hspace{.1in} (B_2\leq 2E_F) 
\nonumber \\
&=&N/2 \hspace {0.4in} (B_2 \geq 2E_F). \label{nb00}
\end{eqnarray}
This should be compared with the BCS theory estimate (Ref. \cite {Blatt}, 
p. 128)
\begin{equation}
N_B(0)\sim (\Delta /E_F)^2 \frac{N}{2} = N(B_2/E_F), \label {bcsnb}
\end{equation}
where here $\Delta$ is the BCS $T=0$ energy gap, and the exact 2D result 
\cite{Miy} $\Delta=\sqrt {2E_FB_2}$ was used in the last step. Since
$N_B \leq N/2$, the estimate implies a breakdown for 
$B_2 \geq E_F/2$ in the BCS case. \\

\smallskip
\noindent {\bf 4. Critical temperature}
\smallskip
\smallskip
    
Neglecting the background unpaired fermions and modeling the entire system as a {\it pure boson gas} of unbreakable CPs but with temperature-dependent boson number density $n_{B}(T)\equiv N_{B}(T)/L^{2}$, the explicit BEC $T_{c}$-formula for 
linear dispersion bosons in 2D \cite {pla} becomes an {\it implicit} one by 
allowing $n_{B}$ to be $T$-dependent, namely 

\begin{equation}
T_{c}=\frac{4\sqrt{3}}{\pi ^{3/2}}\frac{\hbar v_{F}}{k_{B}}\sqrt{n_{B}(T_{c})}.  
\label {(11)}
\end{equation}
This differs from the familiar BEC 3D formula $T_{c}\simeq
3.31\hbar ^{2}n_{B}^{2/3}/m_{B}k_{B}$ for 
quadratic-dispersion bosons. Both equations are special cases of the more 
general expression \cite{pla} of the form $T_c \propto n_B^{s/d}$
for any space dimensionality $d>0$ and any boson 
dispersion relation $\varepsilon_{K} \propto K^{s}$ with $s$\ $>0$. 
%and $C_{s}$ a constant, given by 
%\begin{equation}
%T_{c}=\frac{C_{s}}{k_{B}}\left[ \frac{s\,\Gamma (d/2)\,(2\pi )^{d}n_{B}}{%
%2\pi ^{d/2}\,\Gamma (d/s)g_{d/s}(1)}\right] ^{s/d}  \label{gentc}
%\end{equation}
%where $g_{\sigma}(1)$ is an infinite series that diverges for $\sigma \leq 1$ 
%and is otherwise just the Riemann zeta function $\zeta (\sigma )$ of order 
%$\sigma $.
Solving (\ref {(11)}) with (\ref {n20t26}) and (1) for $K=0$ self-consistently gives $T_{c}/T_{F}$ 
{\it vs.} $B_2/E_F$ as displayed in Figure 3 and compared with empirical values
for cuprates that range \cite {Uem} from $0.01-0.1$.

% FIGURE 3.
\begin{figure}[htbp]
%\hspace{-0.5cm}\epsfig{figure=Tclands.ps,width=8.cm,angle=180}
\hspace{-0.5cm}\epsfig{figure=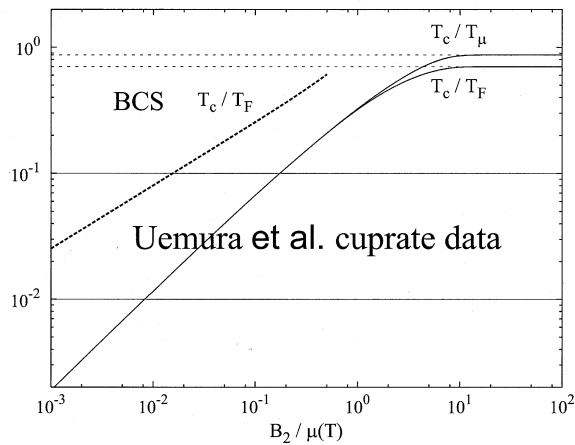,width=8.cm}
%\epsfig{figure=Fig2h.eps,width=8cm}}
%\vspace{.5cm}
\caption[]{Critical BEC temperatures (full curves), for the pure unbreakable-boson gas, in units either of $T_F$ or $T_{\mu} \equiv \mu(T)/k_B$, 
compared with the BCS result (slanted dashed curve), {\it vs.} dimensionless 
coupling $B_2/\mu(T)$. Empirical cuprate data are taken from Ref. \cite {Uem}.} 
\label{Fig 3}
\end{figure}

Also shown 
in the figure are the 
BCS theory $T_c$'s (see also Ref. \cite{DZ}) obtained by solving the  single implicit equation 
\begin{equation}
\int_0^1 \frac{dx}{x} \tanh \frac{T_F}{2T_c}x = \ln \left(\frac{\pi T_c}
{e^{\gamma} B_2}\right), \label{BCS}
\end{equation} 
where $\gamma$ is the Euler constant. Note that $k_B T_c \rightarrow ({e^\gamma/\pi})\sqrt{2B_2 E_F}$ as
coupling goes to zero, and also that $2 \Delta/k_B T_c \rightarrow 2 \pi/e^\gamma \simeq 3.53$. \\

\smallskip
\noindent {\bf 5. BCS and BE condensates}
\smallskip
\smallskip

Finally, Figure 4 depicts in either 2D or 3D both condensates, the BCS one with its {\it single} 
$K=0$ pair-correlation state and the BE condensate \cite {pla} with both (ground) $K=0$ {\it
and} several (excited) $K>0$ CP states that form a \lq\lq band" (shown in the
figure as a discrete spectrum for clarity) extending up to the breakup state 
$K_{0}$ defined by $\Delta_{K_{0}}=0$.  For perfectly linear dispersion CPs, 
i.e., in 2D $\varepsilon _{K}\equiv \Delta _{0}-\Delta _{K}=2\hbar v_{F}K/\pi $, the
breakup CMM wavenumber is then just $K_{0}=\pi \Delta _{0}/2\hbar v_{F}$. As this vanishes with coupling all the excited boson levels collapse  downwards and merge with the ground $K=0$
%\vspace{-2.cm}
% FIGURE 4.
\begin{center}
\begin{figure}[htbp]
\epsfig{figure=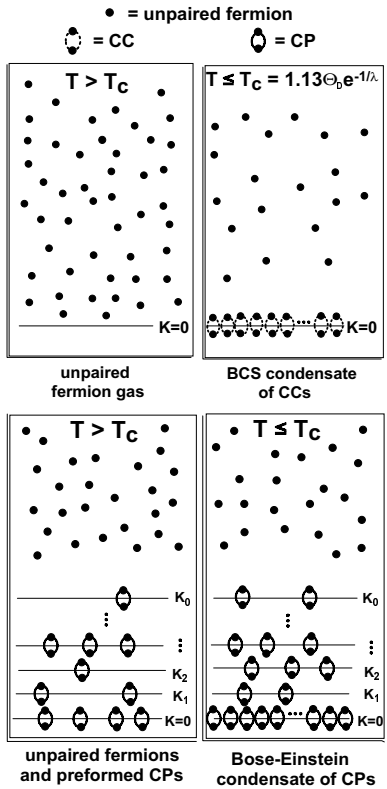,width=8.0cm}
\vspace{0.4cm}
\caption[]{
BCS pair condensate of Cooper correlations (CCs) and BE condensate of CPs, both below $T_{c}$,
compared as explained in text, along with their respective normal
states at $T>T_{c}$.  Horizontal {\it ellipsis} indicate a fractional particle occupation which is
macroscopic, or significant compared with unity.}
\label{Fig 4}
\end{figure}
\end{center}

\noindent level, i.e., the \lq\lq band" shrinks to the 
single ground level. Thus, for zero coupling the BCS condensate appears to be a 
special case of the BE condensate provided that the BCS CCs are essentially CPs, as is widely believed. \\

\smallskip
\noindent {\bf 6. Discussion}
\smallskip
\smallskip

Besides including the background unpaired fer\-mions in the real {\it mixture}
problem with our simple initial s-wave interfermion interaction, further
refinements pending are:  i) realistic Fermi surfaces; ii) Van Hove 
singularities or other means of accounting for periodic-crystalline effects; as 
well as the following interactions iii) the all-important $d$-wave; iv) residual 
interbosonic ones; and v) the crucial CP-fermion interaction vertex. It is precisely the 
latter ingredient that enabled T.D. Lee and coworkers \cite {Lee}, and Tolmachev
\cite {Tolma} more generally, to link BCS and BEC through a relation whereby the BE condensate fraction is proportional to the (BCS-like) fermionic gap
$\Delta(T)$ squared. \\

\smallskip

\noindent {{\bf Acknowledgments.}} Partial support from UNAM-DGAPA-PAPIIT 
(Mexico) \# IN102198, CONACyT
(Mexico) \# 27828 E, DGES (Spain) \# PB98-0124 is gratefully acknowledged.
M.deLl. thanks S.K. Adhikari and V.V. Tolmachev for extensive correspondence.

\end{document}